\documentstyle[12pt,epsfig]{article}
\begin {document}

\begin{center}
{\bf RATIOS OF ANTIPARTICLE/PARTICLE YIELDS IN HADRON-HADRON
AND HEAVY ION COLLISIONS}

Yu.M.Shabelski \footnote{E-mail SHABELSK@THD.PNPI.SPB.RU} \\
Petersburg Nuclear Physics Institute, \\
Gatchina, St. Petersburg 188300 Russia \\
\end{center}
{\it Lecture, given at XXXVII Rencontes de MORIOND, \\
Les Arcs, France, 2002}  \\

\vspace {.5cm}

\begin{abstract}
We compare the numerical results for antibaryon/baryon production ratios in
high energy heavy ion collisions with some theoretical models.
\end{abstract}

\vspace {1.cm}

The first RHIC experimental data collected in [1] show that the values of
antibaryon/baryon production ratios in midrapidity region of Au-Au
central collisions at $\sqrt{s_{NN}} = 130$ GeV are rather small in
comparison with the most of theoretical predictions. Really, the
values of the order of 0.6 for $\bar{p}/p$ and $0.73 \pm 0.03$ for
$\bar{\Lambda}/\Lambda$ were measured [1] whereas the standard Quark-Gluon
String Model (QGSM) [2,3,4,5] predicts in both cases the values more than
0.9 and in the String Fusion Model [1] the predicted values are
about 0.8 and about 0.87, respectevely.

The main part of $B$ and $\bar{B}$ should be produced at high energies as
$B-\bar{B}$ pairs. It means that the additional source of the baryons in the
midrapidity region is needed. The realistic source of the transfer of baryons
charge over long rapidity distances can be realized via the string junction
(SJ) diffusion where it can combain three sea quarks into a secondary baryon
(but not antibaryon), see [6].

In such a picture where the transfere of baryon charge is connected with SJ
exchange there exist three different possibilities to produce a secondary
baryon which are shown in Fig. 1 [7].

This additional transfer of baryon number to the midrapidity region is
governs by contribution [7]
\begin{equation}
G^p_{uu} = G^p_{ud} = a_N \sqrt{z}[v_0 \epsilon (1 - z)^2
+ v_q z^{3/2} (1-z) + v_{qq} z^2] \;,
\end{equation}
where the items proportional to $v_{qq}$, $v_q$ and $v_0$ correspond
to the contributions of diagrams Fig. 1a, 1b and 1c, respectively.
The most important in the midrapidity region at high energies is the
diagram Fig. 1c which obeys the baryon number transfer to rather
large rapidity region.

In agreement with the data the parameter $\epsilon$ in Eq. (1) is rather small.
However different data are in some disagreement with each other, see the
detailed analyses in [7]. Say, $x_F$ -distributions of secondary protons
and antiprotons produced at 100 and 175 GeV/c (triangles, [8]) and at
17.3 GeV/c (squares, [9]) are in better agreement with $\epsilon = 0.05$, as
it is shown in Fig. 2, whereas ISR data [10] are described better with
$\epsilon = 0.2$, see Fig. 3. In both Figs. 2 and 3 the varyants with
$\epsilon = 0.05$ are shown by solid curves and the variants with
$\epsilon = 0.2$ by dashed curves.

Some part of this disagreement can be connected with different energies. In
Fig. 1c two additional mesons $M$ should be produced in one of the strings
that can give the additional smallness [11] at not very high energy.
Another source of disagreement of the data at low and high energies can come
from the fact that the suppression of baryon number transfer to large rapidity
distance $\Delta y$ should be proportional to
\begin{equation}
e^{-(\alpha_{SJ}-1)\cdot  \Delta y} \:,
\end{equation}
and the effective value of $\alpha_{SJ}$ can depend on the energy due to Regge
cut contribution [12].

In the case of $\pi^-p \to \Omega X$ reaction the contribution (1) leads to
the contradiction with additive quark model [13] because the yields of
$\Omega$ in the central region is permanently larger than the yields of
$\bar{\Omega}$.

For the case of hadron-nucleus collisions the yields of secondaries can be
calculated in QGSM by similar way [3, 14] as the $hN$ collisions and also
with including SJ contributions.

\begin{figure}[htb]
\centerline{
\mbox{\epsfig{file=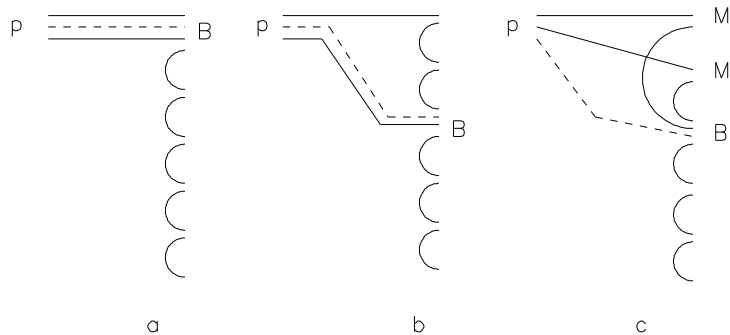,width=0.7\textwidth}}}
\caption{
 Three different possibilities of secondary baryon production in $pp$
collisions: SJ together with two valence and one sea quarks (a), together
with one valence and two sea quarks (b), together with three sea quarks (c).
}
\end{figure}

In the case of heavy ion (A-B) collisions the multiple scattering theory
allows one to account for the contribution af all Glauber-type diagrams
only using some Monte Carlo method where the integrals with dimension of
about $2 \cdot A \cdot B$ should be calculated in coordinate space [15, 16].
The analitical calculations allow one to account only some classes of diagrams
[17, 18, 19]. One of approaches here is so called rigid target approximation
where it is assumed that for the forward hemisphere we can neglect by the
binding of projectile nucleons (i.e. consider them as a beam of free nucleons
and every of them can interact with target nucleus). The last one is considered
as a dense medium. And vice versa, for the backward hemisphere we consider
the target nucleons as a beam of free nucleons and every of them can interact
with dense medium of projectile nucleus.

All details and needed formulae can be found in [5, 19]. The resulting expression for secondary $h$
production in $A-B$ collisions reads as

\begin{figure}[h]
\centerline{
\epsfig{figure=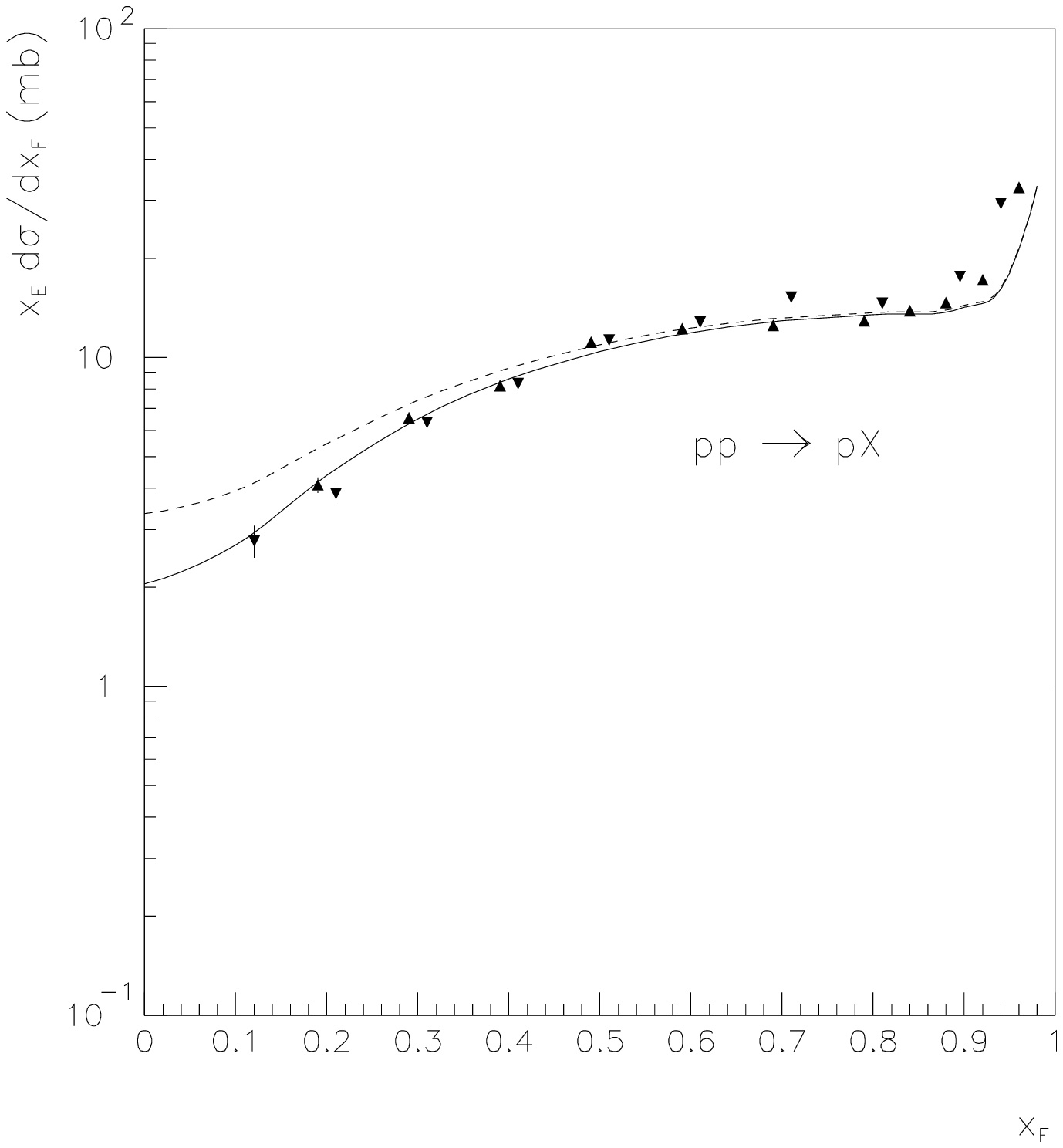,width=0.45\textwidth,clip=}
         \hspace{0.3cm}
\epsfig{figure=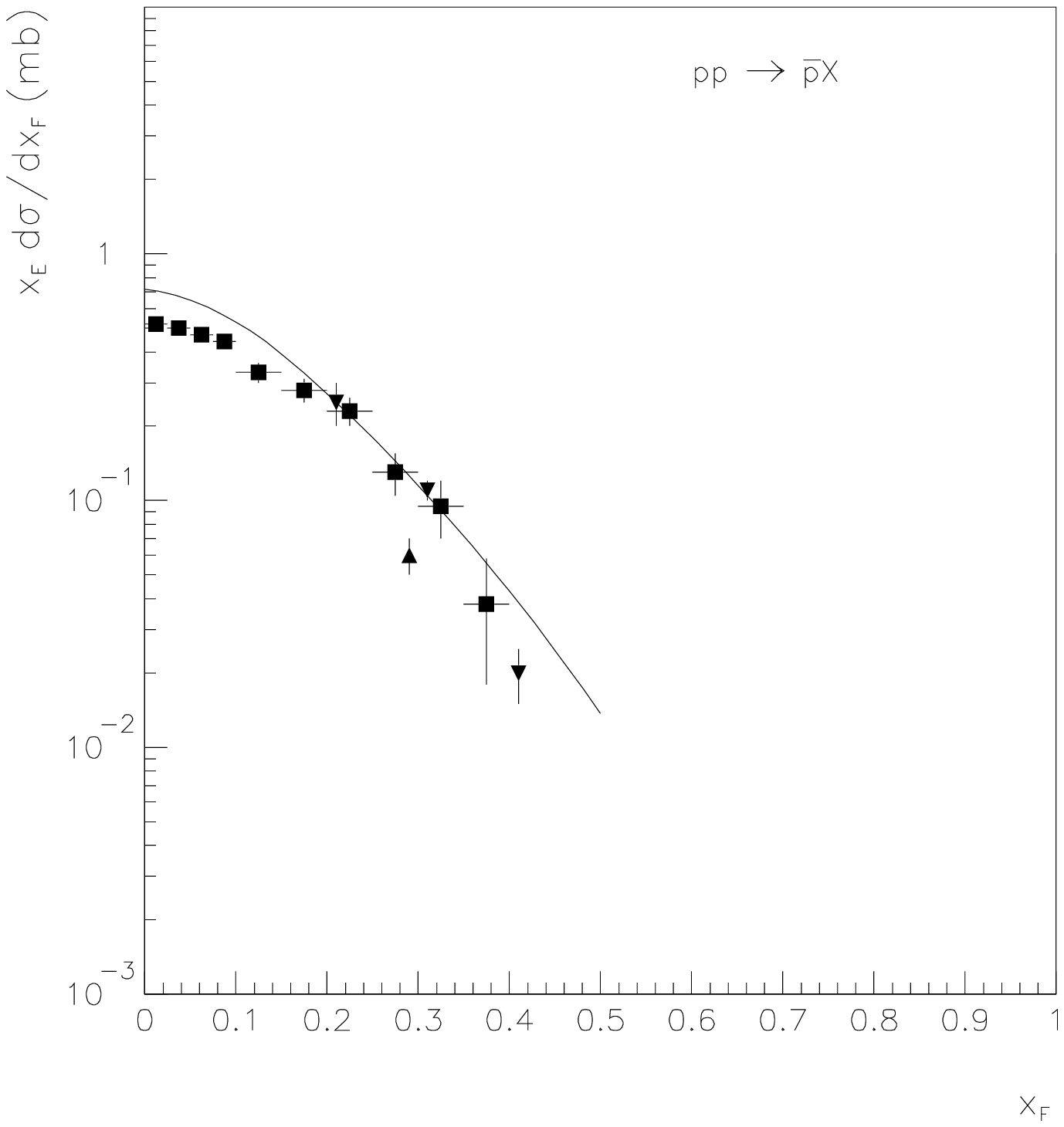,width=0.45\textwidth,clip=}}
\caption{
 $x_F$ -distributions of secondary protons and antiprotons produced
in $pp$ collisions at 100 and 175 GeV/c (triangles, [8]) and at 17.3 GeV/c
(squares, [9]).}
\end{figure}

\begin{figure}[h]
\centerline{
\epsfig{figure=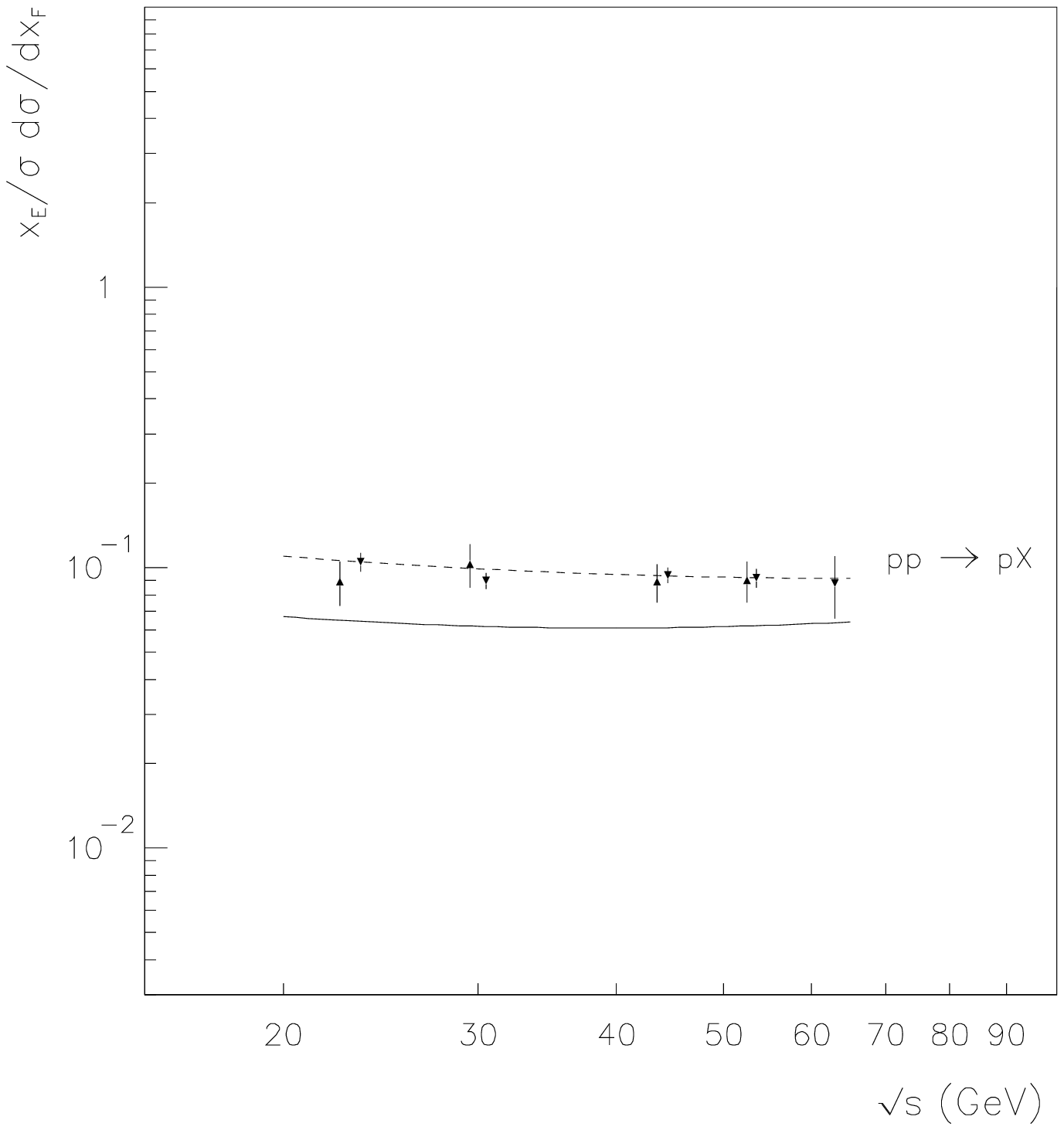,width=0.45\textwidth,clip=}
         \hspace{0.3cm}
\epsfig{figure=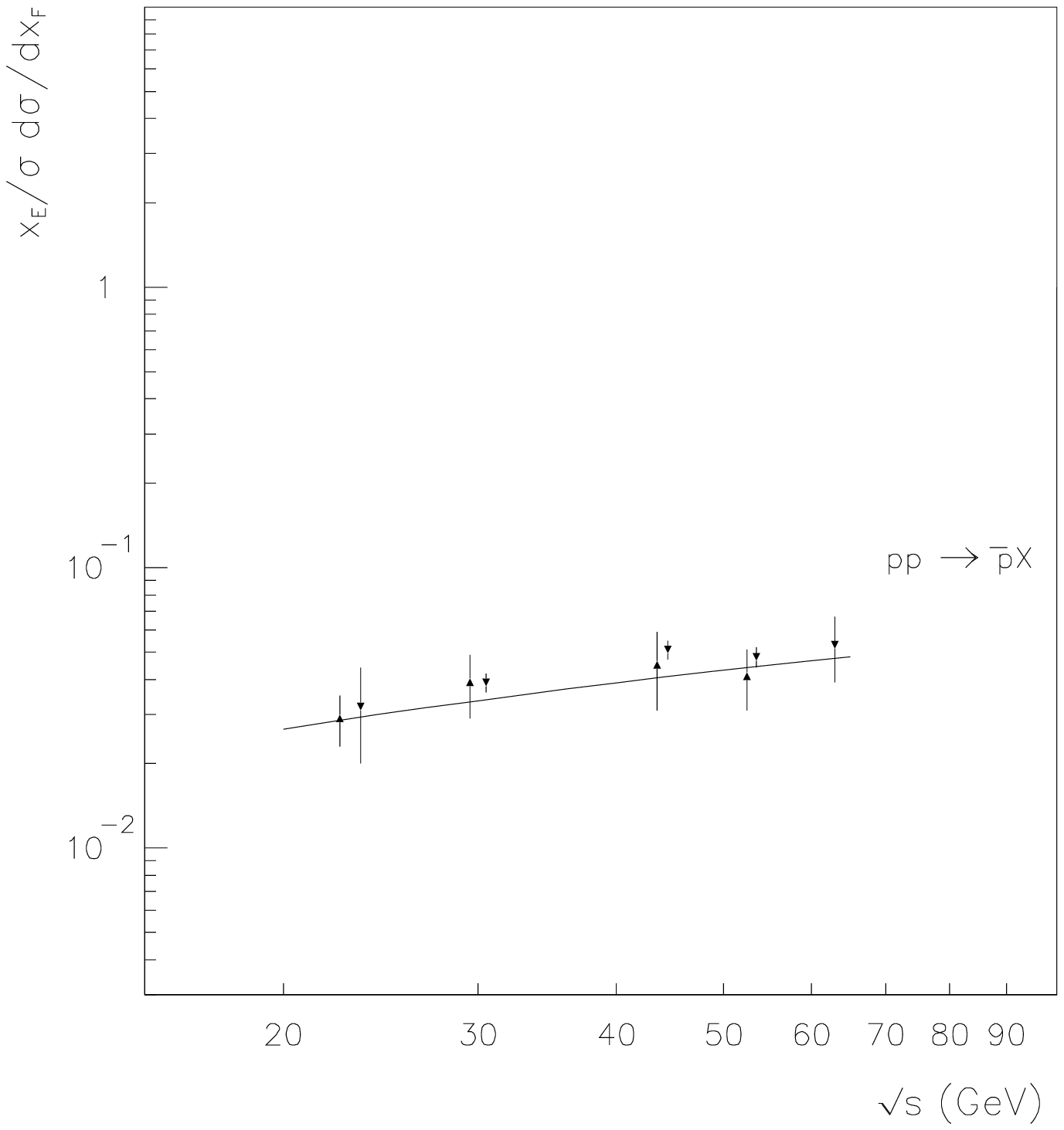,width=0.45\textwidth,clip=}}
\caption{
Secondary proton and antiproton yields at ISR energies
[10] at $90^o$ in c.m.s.}
\end{figure}
\clearpage

\begin{eqnarray}
\frac1{\sigma^{prod}_{AB}} \frac{\sigma(AB \to hX)}{dy} & = &
\theta(y)\langle N_A\rangle \frac1{\sigma^{prod}_{NB}} \frac{\sigma(NB \to
hX)}{dy} + \nonumber\\
& + & \theta(-y)\langle N_B\rangle \frac1{\sigma^{prod}_{NA}} \frac{NA
\to hX}{dy}\ ,
\end{eqnarray}
where $\langle N_A\rangle$ and $\langle
N_B\rangle$ are the average numbers of interacting nucleons in nuclei
$A$ and $B$.  They depend on the $A-B$ impact parameter, $A/B$ ratio,
etc. [21] The presented approach can be consider to be close to some
versionm of intranuclear cascade model.  However, many interference
contributions can be reestablished [22] with the help of AGK cutting
rules [23].

It is necessary to note that at RHIC energies we observed experimentally
rather new effect --- very large shadowing of hadron production due to
the important contribution of diagrams containing the multipomeron
interactions.  This effect is based on the prediction of Gribov theory
for reggeon diagram technique [24] which was consider in details [25,
26] for the case of nucleus-nucleus interactions, and it reproduces the
RHIC data in different model approach [27, 28, 20] based on the Gribov
picture.

\begin{table}
\caption{Antibaryon/baryon yields for secondaries produced at RHIC in
midrapidity region at $\sqrt = 130$ GeV}.
\begin{center}
\begin{tabular}{||c|c|c|c||} \hline \hline
& \multicolumn{2}{c|}{QGSM} &  Exper.  \\  \hline

& $\epsilon = 0.05$ & $\epsilon=0.2$ & \\ \hline

$\bar p/p$ & 0.83  & 0.67  &  $\sim0.6$  \\ \hline

$\bar\Lambda/\Lambda$ & 0.83 & 0.64 & $0.73\pm0.03$ \\
\hline \hline
\end{tabular}
\end{center}
\end{table}

\subsection*{Acknowledgements}
This work was supported by grants NATO PSTCLG 977275 and RFBR 01-02-17095.
I am very grateful to G.Arakelyan, A.Capella, J.Dias de Deus, A.Kaidalov
and R.Ugoccioni for collaboration and useful discussins.

\end{document}